# Crystal Size Improvement of Bi-based Superconducting Whiskers under Stress-controlled Condition

Sayaka Yamamoto[a,b,c], Ryo Matsumoto[a,b], Shintaro Adachi[a], Yoshihiko Takano[a,b],
Hiroyuki Muto[d], and Hiromi Tanaka[c,e]

[a]National Institute for Materials Science, 1-2-1 Sengen, Tsukuba, Ibaraki 305-0047, Japan
[b]Graduate School of Pure and Applied Sciences, University of Tsukuba, 1-1-1 Tennodai, Tsukuba, Ibaraki 305-8577, Japan
[c]National Institute of Technology, Yonago College, 4448 Hikona, Yonago, Tottori 683-8502, Japan
[d]Institute of Liberal Arts and Science, Toyohashi University of Technology, 1-1 Hibarigaoka, Tempaku, Toyohashi, Aichi 441-8580, Japan
[e]Plasma Science and Fusion Center, Massachusetts Institute of Technology, 77 Massachusetts Avenue, NW17, Cambridge, MA 02139, U.S.A.

**Abstract**

Using stress-controlled amorphous precursors, we have successfully grown large $Bi_2Sr_2Ca_{n-1}Cu_nO_y$ (Bi-based) high-transition temperature superconducting whiskers. Especially, under a high compressive stress attained by 1 GPa pelletizing, the crystal size of a whisker was significantly enhanced up to 9.5 mm length (growth period: 96 h) which is 2.4 times larger than that of the conventional amorphous-precursors method. In addition, by using the stress-controlled precursors, the number of the obtained Bi-based whiskers were also increased by a factor of 1.8.

## 1. Introduction

Bi-based high-transition temperature ($T_c$) superconducting whisker (hereafter, Bi-based whisker) is one of the most promising materials for the application of high-electric-current equipment such as lead wires and electric power cables since the whisker shows high critical current density ($J_c$) above $10^5$ A/cm beyond a criterion for a practical use, compared to that of bulk sample[1-3]. Well-established growth method for the Bi-based whisker is an $Al_2O_3$-seeded glassy quenched platelets (ASGQP) process which typically provides 10 mm length of the whiskers[4,5]. It is quite important to increase the crystal size of Bi-based whisker for practical use, for example, an application for superconducting wire. To improve the crystal size, it is necessary to understand the driving force of the growth and to increase it for crystal growth of Bi-based whiskers.

The growth mechanism for the metal whiskers, for example, Zn whiskers and Sn whiskers have long been studied because they sometimes cause an electrical accident in an electrical circuit. According to the previous study, the whiskers are rapidly grown from electroplate under compressive stress condition[6,7]. In the growth models for the metal whiskers, the compressive stress accelerates the diffusion of composition from the precursors to whiskers [8]. In the growth of Bi-based whiskers, it has been previously implied that the compressive stress possibly acts as one of the driving forces for the growth, although it has not yet been evaluated quantitatively[9-11]. If the compressive stress in the precursors can be controlled, it is expected to improve the crystal size of Bi-based whiskers.

In this study, we considered the growth mechanism for improving the crystal size of Bi-based whiskers. As a first step, the relationship between the whisker length and the curvature of the precursor was quantitatively investigated by changing the quenching pressure for precursor synthesis preparations. Here, the curvature is considered to relate the inner stress of precursors. In addition, we demonstrated an improvement of the crystal size of Bi-based whiskers under stress-controlled condition. To change the compressive stress in the amorphous precursor, the grassy quenched precursors were grounded into rough powders, and then pelletized by various pressures. The superconducting properties of $T_c$ and $J_c$ ware evaluated via electrical measurements using a four-probe method using gold wires and silver-paste painting.

## 2. Experimental
### 2.1 Growth of Bi-based whiskers by ASGQP method

The precursor of Bi-based whisker was prepared by typical ASGQP method [4, 5, 12, 13] as show in the schematic images of Figure 1(a). Starting materials of $Bi_2O_3$, $SrCO_3$, $CaCO_3$ and CuO were mixed in a nominal molar ratio of Bi : Sr : Ca : Cu = 2 : 2 : 2 : 4. The mixture was heated at 1200 ºC for 0.5 h in the air. The completely molten mixture was poured on to the iron plate where $Al_2O_3$ powders were scattered, and then rapidly pressed by another iron plate. Although the quenching process in the conventional ASGQP process is not quantitative and strongly depends on human power [4, 5, 12, 13]. In this study, we evaluated the quenching pressure quantitatively and prepared the various precursors with pressures of 0.82, 10.0, and 60.9 MPa, as shown in Figure 1(b). The $Al_2O_3$ powders which are a catalyst for the whisker growth are incorporated into near the surface of the precursors [4, 14].

According to the recent work [15], the pulverized amorphous precursors promote the whisker growth. As shown in Figure 1(c), we also ground the precursors into rough powders of which the particle size is about 1.0 × 1.0 × 0.3 mm³. Bi-based whiskers were grown by annealing the precursors in a tube furnace with following conditions; an oxygen gas-flow of 120 ml/min, a growth temperature of 870 - 890 ºC, and a growth period of 24 - 96 h. To investigate the relationship between the crystal size of Bi-based whiskers and the compressive stress in the precursors quantitatively, we measured a curvature of the precursors after the whisker growth. The curvature was defined by the inversion of curvature radius. The curvature radius was estimated by a scanning electron microscopy (SEM) observation using TM3000 (Hitachi High-Technologies).

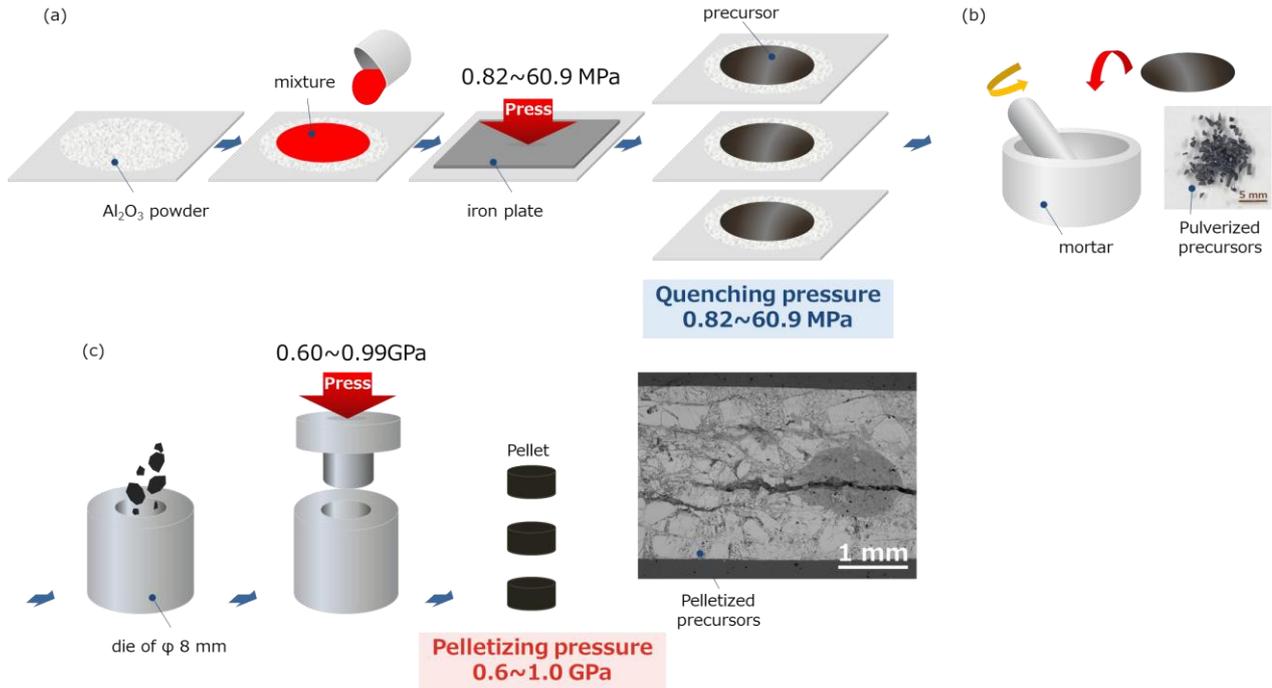

**Figure 1. Schematic images for a process of sample preparation by various methods. (a) conventional ASGQP method [4, 5, 12, 13], (b) Pulverized precursor method [15], and (c) Newly proposed stress-controlled method.**

## 2.2 Growth of Bi-based whiskers from stress-controlled precursors

To control the compressive stress in the precursors, we pelletized the crashed amorphous precursors with various pressures as shown in Fig. 1(c). The precursors with the size of about 1.0 × 1.0 × 0.3 mm³ were prepared with the compositional ratio of Bi : Sr : Ca : Cu : Al = 2 : 2 : 2 : 4 : 0.75, referring to the previous report[5]. The prepared rough powders were blended with much more fine powders with size of around several microns to decrease a porosity of the pellet. The powders of the precursor were pelletized in a die with a diameter of φ 8 mm under the pressures of 0.6, 0.8, and 1.0 GPa. From cross-sectional SEM image of pelletized precursors as shown in fig. 1(c), the various size of the precursors are packed tightly. The pellets were annealed under the same condition for the growth of Bi-based whiskers as described in the section 2.1. The superconducting properties, which are $T_c$ and $J_c$ of the obtained Bi-based whiskers were evaluated by resistivity-temperature ($\rho$-$T$) and current-voltage (*I-V*) measurements using a four-probe method.

## 3. Results and discussion
### 3.1 Relationship between a curvature of heated amorphous precursor and crystal sizes of whiskers

Figure 2 shows SEM images of the precursors after growing Bi-based whiskers by pulverized precursor method with quenching pressure of (a) 0.82 MPa and (b) 60.9 MPa. The growth period of the Bi-based whiskers was 24 h. As show in the dashed line indicating the curvature of the precursors, the highly compressed precursor by 60.9 MPa exhibits larger curvature. Moreover, the crystal size along a longitudinal direction of whiskers obtained from the more bended precursors was longer than that of the other one.

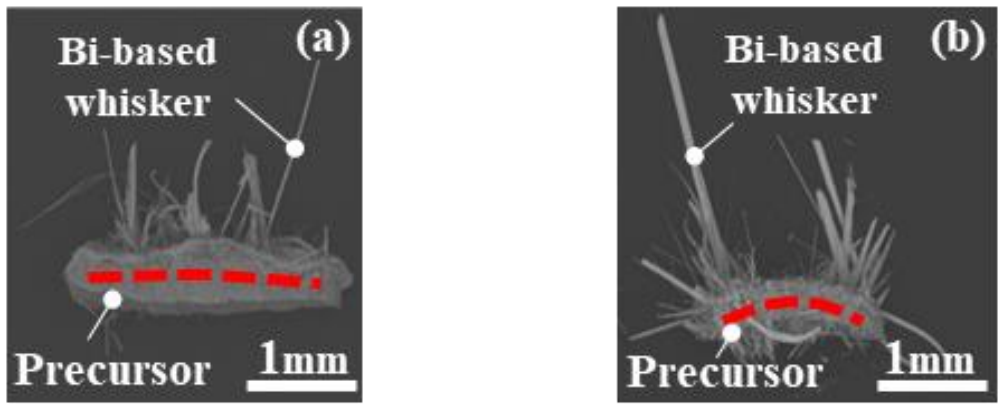

**Figure 2.** SEM images for Bi-based whiskers from precursors quenched by (a) 0.82 MPa and (b) 60.9 MPa. The dashed lines are guide for eyes about the curvature.

To investigate a relationship between the curvature of the precursor after the whisker growth and the crystal size of the obtained Bi-based whiskers quantitatively, we prepared various precursors by changing the quench pressures and grew the Bi-based whiskers. Figure 3 shows the maximum crystal sizes of the obtained Bi-based whiskers as a function of the curvature of the precursors after the whisker growth. The curvature of the precursors after the whisker growth increase with an increase of the quenching pressure in the ASGQP process. This result indicates the compressive stress into the precursor is controllable by changing the quenching pressure. The crystal sizes of the grown whiskers tend to increase linearly with the increase of the curvature of the precursors as shown in a guide for eyes of dashed line derived by least square method. Especially, the bended precursors with the curvature of 1.4 mm$^{-1}$ prepared by 60.9 MPa provided the Bi-based whiskers with the maximum crystal size of 3.7 mm. The deformation of the precursor could be considered a release of the compressive stress which is depending on the quenching pressure. According to the positive correlation between the curvature of the precursors and crystal sizes of the whiskers, the compressive stress in the precursors could be considered as one of the driving forces for the growth of the Bi-based whiskers.

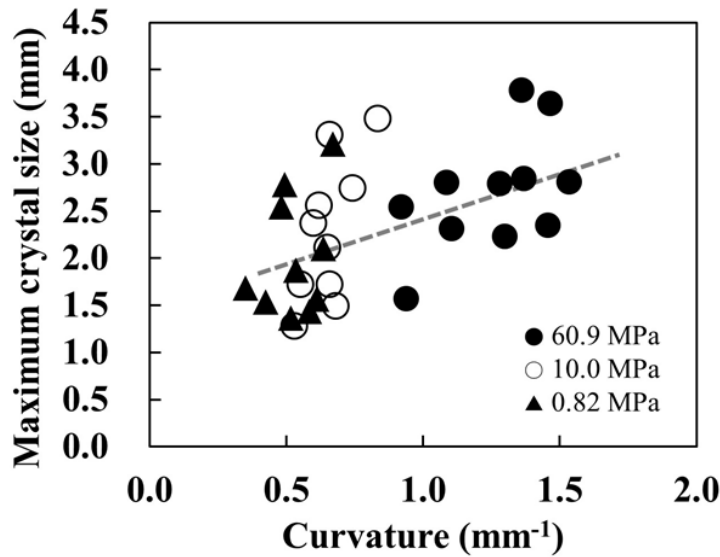

**Figure 3.** The relationship between the curvature of precursors and the maximum crystal sizes of Bi-based whiskers. The quenching pressure of the precursors were changed from 0.82 to 60.9 MPa. The growth period was 24h. The dashed line is guide for eyes.

**3.2 Whisker growth under compressive stress-controlled condition**

To enhance the whisker growth effectively using the driving force of compressive stress, we prepared compressive stress-controlled precursors by pelletizing the pulverized precursors under various pressures. Figure 4 shows an optical image of the typical whiskers grown from the pelletized precursors. The whiskers were grown from the entire region of pellet surface with high density.

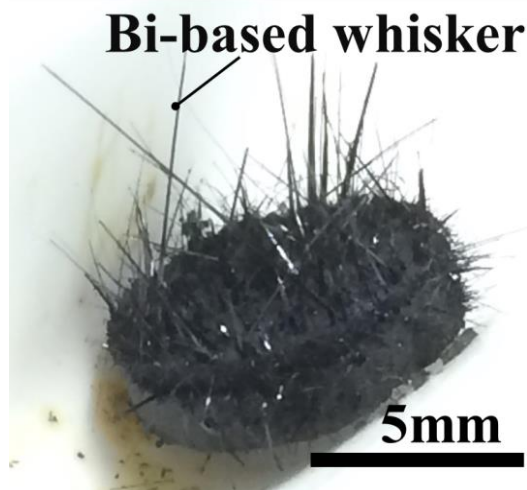

**Figure 4.** Optical image of Bi-based whiskers grown from a compressive stress-controlled glassy quenched precursor.

The square symbols of fig. 5 (a) shows the pelletizing pressure dependence on the maximum crystal size of Bi-based whiskers obtained with a growth period of 24 h. The crystal size of the grown whisker linearly increased as a function of the pelletizing pressure, and reached 5 mm maximally when the pressure becomes 1.0 GPa. We additionally confirmed the stress-controlling effects for the precursors prepared without $Al_2O_3$ scattering on the iron plate in the quenching process to eliminate the effects of $Al_2O_3$ catalyzer. As shown in the circle symbols in the fig. 5(a), the maximum size of the Bi-based whiskers linearly increased against the pelletizing pressures even without the $Al_2O_3$ catalyzer. It could be considered that the compressive stress in the precursors works effectively necessary condition for the whisker growth and controllable by changing the applied pelletizing pressure. The scatter of $Al_2O_3$ catalyst can promote the whisker growth by combining the compress pressure effect.

Figure 5 (b) shows a comparison for the maximum crystal sizes of Bi-based whiskers obtained by the conventional pulverized-precursor method using different grain size of large (~4×4 $mm^2$) and small (~1×1 $mm^2$) precursors, and compressive stress-controlled method using the pelletizing process. The maximum crystal sizes of Bi-based whiskers grown by conventional method were 4.0 mm and 7.5 mm length in using large and small precursors, respectively. The improvement of the crystal size will be originated from the difference of the compressive stress in the precursors. Moreover, the crystal size significantly increased up to 9.5 mm length by using the stress-controlled method, which length is longer by 2.4 times as compared with the conventional method with large precursor.

By using the stress-controlled precursors for the Bi-based whisker growth, not only the length but also the efficiency of the growth for Bi-based whiskers from the same amount of a precursor was improved. Figure 5 (c) shows the number of Bi-based whiskers above 1 mm length grown from 1 g of the precursor prepared by conventional ASGQP method and the stress-controlled method with/without $Al_2O_3$ catalyst. The conventional method provided 15 whiskers / g. In contrast, the number of the obtained whiskers were enhanced up to 26 whiskers / g from pressure-controlled precursors. Even on the stress-controlled precursors without $Al_2O_3$ catalyzer, 14 whiskers / g were grown which is comparable with that of the conventional method. Therefore, the compressive stress is effective to drive the crystal growth of the Bi-based whiskers.

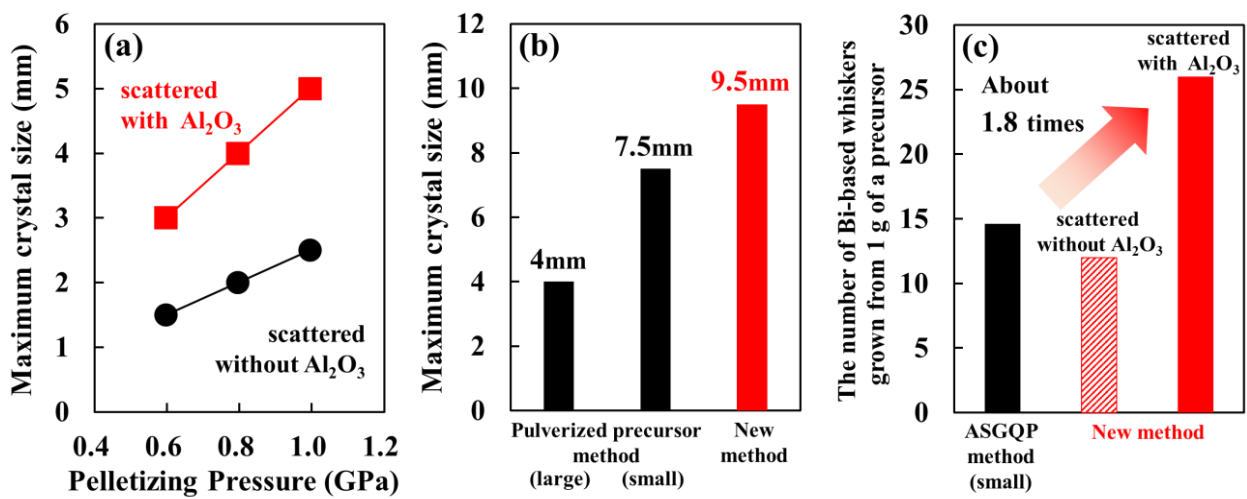

**Figure 5.** (a) The pelletizing pressure dependence of the maximum crystal size of the Bi-based whiskers. (b) The comparison of the maximum crystal size of Bi-based whiskers grown by various growth methods. (c) The comparison of the number of Bi-based whiskers grown from 1 g of a precursor on various growth conditions.

We demonstrated the electrical transport measurements of the obtained whiskers to confirm the superconducting properties of $T_c$ and $J_c$. Figure 6 (a) shows a temperature dependence of a resistivity for the obtained Bi-based whiskers from the pelletized precursors by 1.0 GPa (growth period: 96 h). The resistivity was linearly decreased from room temperature with a decrease of the temperature, and rapidly dropped at 109.3K, corresponding to the superconducting transition from an intergrowth of $Bi_2Sr_2Ca_2Cu_3O_{10+\delta}$ (Bi-2223) phase [3, 12, 13, 16]. At 77.3 K the resistivity completely dropped to zero in good agreement with a superconducting transition of main phase of $Bi_2Sr_2CaCu_2O_{8+\delta}$ (Bi-2212) [5, 12, 13]. Figure 6 (b) shows $I$-$V$ curve of the whisker with cross-sectional area of 28 $\mu m^2$. The voltage appeared at 40.4 mA, namely the $J_c$ value of the whisker was $1.44 \times 10^5 A/cm^2$ at 10K in self-field. These $T_c$ and $J_c$ values and the intergrowth feature of Bi-2223 are consistent with previous reports for the growth of Bi-based whiskers [1, 12, 17, 18].

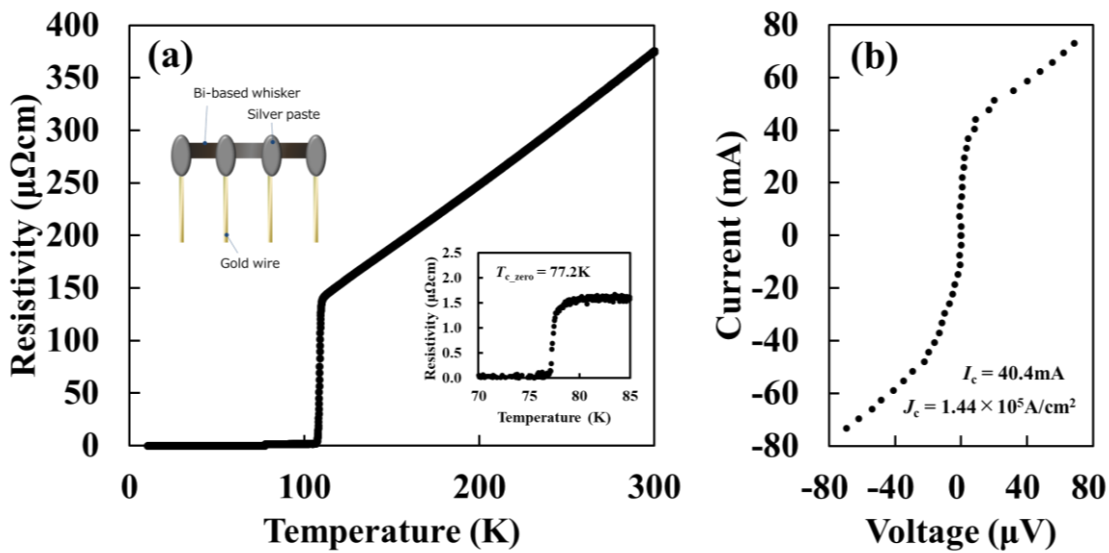

**Figure 6. The electrical transport properties for the Bi-based whiskers grown from the stress-controlled precursors of (a) $\rho$-$T$ characteristics and (b) $I$-$V$ characteristics.**

4. Conclusion

In this study, we focused on the relationship between the curvature of precursors and crystal size of grown whiskers to reveal a driving force of whisker growth. According to the results, more largely bended precursors tended to provide longer Bi-based whiskers, indicating a contribution of compressive stress for the whisker growth. To promote the whisker growth adopting the driving force of compressive stress, we grew the Bi-based whiskers from compressive stress-controlled precursors. By using this method, we succeeded in increasing the maximum crystal size of Bi-based whiskers by about 2.4 times compared to conventional ASGQP method. From these results, we conclude that the compressive stress in the precursors can be one of the driving forces for Bi-based whisker and the crystal size can be increased by increase of the stress. This is important suggestion to clarify the growth mechanism of whiskers not only Bi-based but also the other whiskers.

**Acknowledgment**

The authors thank Mr. S. Tsunashima, Mr. S. Tanaka and Mr. N. Kataoka for discussions. This work was partially supported by JSPS KAKENHI Grant Number JP17K06362 and JP26820119, JP17J05926 and JP19H02177, and JST CREST Grant No. JPMJCR16Q6, JST-Mirai Program JPMJMI17A2.